# Analytical approach of Brillouin Amplification over threshold

**FIKRI SERDAR GÖKHAN,**[1,*] **HASAN GÖKTAŞ,**[2-3] **VOLKER J. SORGER**[2]

[1]*Department of Electrical and Electronic Engineering, Alanya Alaaddin Keykubat University, Kestel, Alanya, Antalya, Turkey*
[2]*Department of Electrical and Computer Engineering, The George Washington University, Washington,D.C. 20052, USA*
[3]*Department of Electrical and Electronic Engineering, Harran University, Sanliurfa, 6300, Turkey*
*\*Corresponding author: fsgokhan@gmail.com*

**We report on an accurate closed-form analytical model for the gain of a Brillouin fiber amplifier that accounts for material loss in the depleted pump regime. We determined the operational model limits with respect to its relevant parameters and pump regimes through both numerical and experimental validation. As such, our results enable accurate performance prediction of Brillouin fiber amplifiers operating in the weak pump, high–gain, and saturation regimes alike.**

## 1. INTRODUCTION

Stimulated Brillouin scattering (SBS) is the most efficient nonlinear amplification mechanism in optical fibers, in which a large gain may be obtained under the pump power of several milliwatts [1]. This has led to the design of Brillouin Fiber Amplifier (BFA) and has been implemented in a wide range of applications, such as active filters due to its narrowband amplification feature [2], or in the control of pulse propagation in optical fibers [3]. The BFA can also be used to measure strain and temperature [4], which has led to the design of distributed Brillouin sensors (DBS). Here strain and temperature can be measured along the whole fiber length. The performance of the BFA fiber sensors needs to account for pump depletion which can be investigated using the solution to the system of coupled intensity equations which describe the interaction between the pump and the Stokes wave due to SBS. The solution of the mentioned ordinary differential equations (ODEs) is a two-point boundary value problem comprising the initial values of the pump and the Stokes waves at different input points. Such nonlinear ODEs are typically solved numerically due to their inherent complexity. The analytical expression to the boundary value problem of these ODEs, thus far, was only introduced for lossless media, with the exception of an analytical solution of integration constant $C$ [5]. However, this solution is not applicable to BFAs where wave interaction occurs over tens of kilometers and loss is a significant effect, as investigated here. Furthermore, loss-induced pump depletion adversely affects the accuracy of the aforementioned sensors [6]. Therefore, an analytical solution of ODEs must include both the attenuation factor as well as depletion effect.

Several attempts have been undertaken to find a general analytical solution to SBS equations in a lossy medium [7-9]. In [7], the system of ODEs is reduced to a single equation, which can only be integrated numerically. In [8], the proposed solution results in a system of two transcendental equations to be solved numerically not allowing for a closed-form solution. Below critical pump powers, the undepleted pump approximation (UPA) can be used. However, DBS applications typically require pump powers above the SBS threshold and in this region the pump becomes depleted and UPA strongly overestimates the real Brillouin gain, which renders this solution unusable for regimes above the critical pump threshold [9].

In another approach [10], a perturbation technique is successfully applied to obtain an analytical solution of pump and probe evolution for a lossy fiber using the solution obtained for the lossless fiber. This solution has some bottlenecks, however; the conserved quantity solution of lossless fiber, underestimates the real solution especially (i) when the Stokes pump power is about miliwatts (saturation region), and (ii) when the fiber length is increased beyond 20 km where the imaginary permittivity becomes a loss factor impacting the complete solution. On the other hand, introducing simplifications to the perturbation parameters leads to inaccurate representations.

Recently, an approximate analytic solution of the steady state coupled equations has been proposed [11]. However, these equations include the term for the Stokes intensity at the fiber input, which is not known prior to the computation. In Ref.[12], analytical solutions were obtained in fiber lengths of up to a few kilometers by neglecting the attenuating terms due to the short fiber lengths inherent in the model, which limits sensor applicability which can be fifty to hundred kilometers. As such the problem of accurately treating losses in SBS is yet outstanding.

Here we modified the perturbation method applied from Ref. [10] and extensively improved both the pump and Stokes wave analytical equations for any fiber length, attenuation, pump and stokes powers. The analytical solution is divided into two parts; the first one is the high gain region, where the depletion is moderate applicable to low Stokes powers up to 1-2mW. The second part describes the saturation region where high Stokes powers (>1-2 mW) are used and depletion is high, i.e., when low-gain amplification occurs. Following this formulism, we derive an approximation for the Brillouin gain of the BFA analytically. The pump condition between the regions where high- and low gain amplification occurs is derived. The proposed solution is validated both experimentally and numerically, where we consider a full integration of coupled ODEs. Lastly, we discuss the range of validity ranges of the found solution

## 2. Theoretical model

The coupled ODEs for the evolution of the intensities of pump $I_p$ and Stokes $I_s$ can be written as [1,15]

$$dI_p / dz = -g_B I_p I_s - \alpha I_p$$
$$dI_s / dz = -g_B I_p I_s + \alpha I_s \quad (1)$$

here $0 \leq z \leq L$ is the propagation distance along the optical fiber of the total length $L$, $\alpha$ is the fiber loss coefficient, $g_B$ is the Brillouin gain coefficient. Note that unlike in the calculation of the SBS threshold when the Stokes wave is initiated by noise [5], here we assume a Stokes wave launched from the rear end of the fiber. Then the known values of the input pump intensity $I_{p0}$ and the input Stokes intensity $I_{sL}$ are the boundary values.

In our approach, we represent the unknowns through perturbations to the lossless systems, i.e., as

$$I_{p(1,2)} = I_{pp}\left(1-\rho_{(1,2)}\right), \quad I_{s(1,2)} = I_{ss}\left(1-\mu_{(1,2)}\right) \quad (2)$$

where $I_{pp}$ and $I_{ss}$ are the solutions of the lossless system, i.e.,

$$dI_{pp}/dz = -g_B I_{pp} I_{ss}, \quad dI_{ss}/dz = -g_B I_{pp} I_{ss} \quad (3)$$

with the same boundary values $I_{pp}(0) = I_{p0}$, $I_{ss}(L) = I_{sL}$ as in Eqs.(1). The solution of Eq.(3) is [5]:

$$I_{pp}(z) = c_{1,2} I_{p0} \cdot \left[I_{p0} + (c_{1,2} - I_{p0})\exp(-c_{1,2} g_B z)\right]^{-1}$$
$$I_{ss}(z) = c_{1,2}(I_{p0} - c_{1,2}) \cdot \left[I_{p0}\exp(c_{1,2} g_B z) + (c_{1,2} - I_{p0})\right]^{-1} \quad (4)$$

Where $c_{1,2} = I_{pp} - I_{ss}$ is the conserved quantity of Eq. (3). To explicitly find the value of parameter $c_{1,2}$ we approximately solve the equation;

$$I_{sL} = c_{1,2}(I_{p0} - c_{1,2}) \cdot \left[I_{p0}\exp(c_{1,2} g_B L) + (c_{1,2} - I_{p0})\right]^{-1} \quad (5)$$

using boundary condition $I_{sL}$. The solution of $c_{1,2}$ differs for high-gain and saturation region. We have obtained $c_{1,2}$ by eliminating different parameters of second order differential equations.

### A. Solution for High-Gain Region

In this region, the solution of $c_{1,2}$ becomes,

$$c_1 \simeq \frac{1}{k} \cdot \left( \begin{array}{c} \Lambda + \ln\left(\Lambda\left(1-\frac{\Lambda}{k}\right)\right) - \ln\left(1-\frac{1}{e^\Lambda}\right) \\ -\ln\left(1+\frac{\Lambda}{k(e^\Lambda-1)}\right) \end{array} \right) \cdot I_{p0} \quad (6)$$

where $\varepsilon = I_{sL}/I_{p0}$, $k = g_B I_{p0} L$, $\Lambda = -\ln(\varepsilon k)$. From Eqs. (1)-(3) we then obtain a system of ODEs for the perturbation $\rho$ and $\mu$:

$$\frac{d}{dz}\rho_1 = (-g_B I_{ss}\mu_1 + \alpha)(1-\rho_1)$$
$$\frac{d}{dz}\mu_1 = -(g_B I_{pp}\rho_1 + \alpha)(1-\mu_1) \quad (7)$$

with the boundary values $\rho_1(0) = 0$ and $\mu_1(L) = 0$. In this region $c_1$ is used in the calculation of $I_{pp}$ and $I_{ss}$ and solution of Eq.(7) leads

$$\rho_1(z) \simeq 1 - e^{\frac{\alpha z(c_1 g_B + \alpha)}{c_1 g_B - \alpha}} \times \left(\frac{I_{p0}\alpha e^{(c_1 g_B z \times \Phi)} - (c_1 - I_{p0})(c_1 g_B - \alpha)}{-c_1\left[(c_1 - I_{p0})g_B - \alpha\right]}\right)^{\frac{-2\alpha}{\Phi(c_1 g_B - \alpha)}} \quad (8)$$

where
$$\Phi = (0.1 \times \log_{10}(P_{sL}) + 1.7)$$
and

$$\mu_1(z) \simeq -\frac{0.91\Lambda e^{-c_1 g_B z} e^{1.1z(c_1 g_B + \alpha)} e^{-1.1\alpha L}\left(1-e^{1.1\alpha(L-z)}\right)}{(I_{p0} - c_1)^{0.1}} \quad (9)$$

The evolution of Pump and Stokes waves for High-Gain region is;

$$I_{p(High-Gain)} = I_{pp}(1-\rho_1) \quad (10a)$$

$$I_{s(High-Gain)} = I_{ss}(1-\mu_1) \quad (10b)$$

The UPA solution for the $I_p(z)$ and $I_s(z)$ is;

$$I_p(z) = I_{p0}e^{-\alpha z}$$
$$I_s(z) = I_{sL}e^{\frac{I_{p0}g_B}{\alpha}e^{-\alpha z}\left(1-e^{-\alpha(L-z)}\right)-\alpha(L-z)} \quad (11a)$$

The BFA gain is defined as the ratio of the amplified Stokes power and the launched Stokes power $G_{BFA}=P_S(0)/P_S(L)$.

$$G_{BFA} = e^{I_{p0}g_B \frac{(1-e^{-\alpha L})}{\alpha} - \alpha L} \quad (11b)$$

The parameters in Fig. 1 are chosen to model a 10 km long BFA based on a standard single-mode fiber ($\alpha$ =0.215 dB/km) with the pump power of either 10 mW ($k$=14) or 50 mW ($k$=70). As can be seen from Fig. 1, approximation (10a) accurately describes depletion of the transmitted pump not only for moderate pump powers that are the SBS threshold [Fig. 1(c)], but also for strong input pump and Stokes powers extending saturation region [Fig. 1(d)]. The prediction of the analytical formula for the amplified Stokes wave (10b) is quite accurate even for high pump powers [Fig. 1(b)]. Note that a discrepancy in the initial value of the Stokes wave $I_s$ ($z=L$) [Fig. 1(b)] is due to the approximation taken in Eq.(6).

Stokes wave ($P_{s0}$) at the pump input [Fig.2]. Stokes and pump sources are Erbium doped fiber lasers, whose linewidths are smaller than 100 kHz. Their frequency difference is controlled with a phase locked loop and is locked to the Brillouin frequency, $\nu_B$, of the fiber under test. Power meters are used to monitor the input pump power $P_{p0}$, the transmitted pump power $P_{pL}$, the launched Stokes power $P_{sL}$, and the amplified Stokes power $P_{s0}$. Two fibers with lengths of 10 km. were experimentally studied.

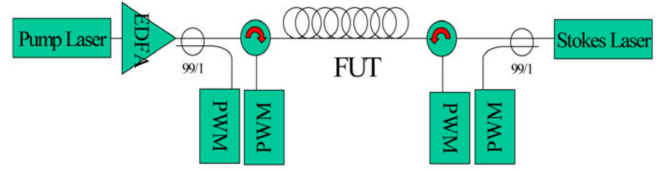

Fig. 2. Schematic of the Brillouin amplifier configuration. $P_{p0}$ and $P_{pL}$ input and output pump power respectively, while $P_{s0}$ and $P_{sL}$ are output and input Stokes power respectively. Pump and Stokes Lasers are Erbium doped Fiber lasers. PWM: power meter; FUT: fiber under test. $\lambda_{pump}$ = 1549.5 nm.

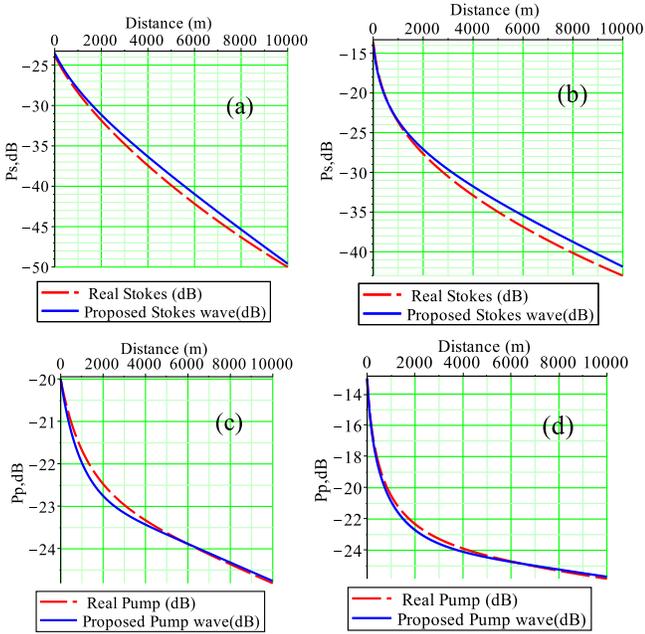

Fig. 1. Stokes wave $P_s$ (a), (b) and Pump wave $P_p$ (c),(d) as a function of fiber distance, $z$. In [(a), (c)], $P_{p0}$=10 mW, $P_{sL}$= 10µW, ($k$ = 14), in [(b), (d)] and $P_{p0}$=50 mW, $P_{sL}$= 50µW ,($k$ = 70). Thick solid curves, approximate analytical solution (10a) and (10b); dashed curves, full numerical solution of Eq. (1). In both figures, L=10 km, $A_{eff}$ = 80µm², $g_B$=1.091x10⁻¹¹, $\alpha$=0.215 dB/km and $\varepsilon$=10⁻³.

The BFA gain can be calculated from the approximation in (10b), Eq.(4), and expressed in terms of the physical parameters as

$$G_{BFA} = \frac{A_{eff}}{P_{sL}}\left(\frac{P_{p0}}{A_{eff}} - c_1\right) \cdot \left(1 - 0.91 \frac{\ln\left(\frac{P_{sL}}{A_{eff}}\right) g_B L \left(e^{-1.1\alpha L} - 1\right)}{\left(\frac{P_{p0}}{A_{eff}} - c_1\right)^{0.1}}\right) \quad (12)$$

To verify Eq. (12) we performed measurements in a standard single-mode optical fiber setup, where we measure the output

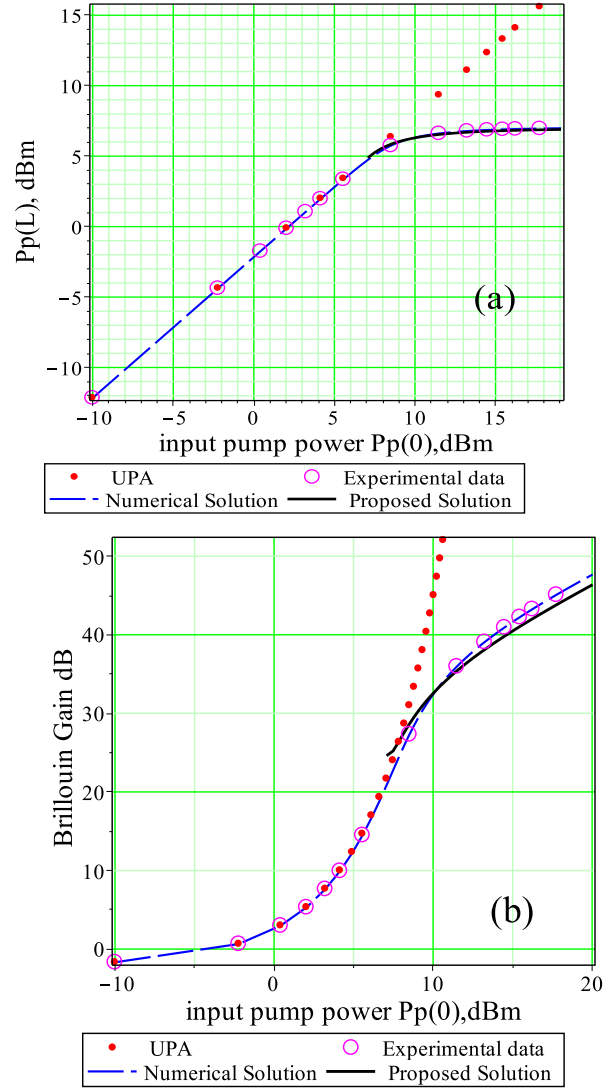

Fig.3. (a) Transmitted pump power $P_p(L)$ and (b) BFA gain $G_{BFA}$ versus the input pump power $P_p(0)$. Open circles, experimental data; thick solid curves, predictions of the analytical formula (12); solid circle curves, calculations based on UPA; dashed curves, numerical results. In both figures, L=10 km, $P_{SL}$ =−1.55 µW, $A_{eff}$ = 80µm², $g_B$=1.091x10⁻¹¹, α=0.215 dB/km.

The experimental results are plotted in [Fig. 3] together with theoretical predictions. We find a high-level of agreement for the transmitted pump for all input pump levels above critical pump power [Fig. 3(a)]. As for the amplifier gain $G_{BFA}$, we find an excellent agreement between predictions from our analytical formula (12) and the measured gain [Fig. 3(b)]. We note, that Eq. (12) is only applicable when the pump power exceeds the critical value, $P_{p0} > P_{cr} \approx \left( \Lambda + \frac{A_{eff}}{0.08 \times g_B L} \Lambda \right)$. For the weak-pump region, the UPA-based estimation for the Brillouin gain can be used [Fig. 3(b)]. The accuracy of Eq. (12) remains high for any fiber length even it extends more than 100 km. The trend in [Fig. 3(b)] remains same for any fiber length.

**B. Solution for Saturation Region**

If $\Lambda < 0$, the saturation region dominates showing high nonlinearity. In this region the solution of Eq.(6) can be analytically defined as,

$$c_2 \simeq -0.99 \frac{P_{sL}}{A_{eff}} - \frac{1}{g_B \cdot L} \text{LambertW}\left( -\frac{P_{sL}}{A_{eff}} g_B L \cdot e^{-0.99 \frac{P_{sL}}{A_{eff}} g_B L} \right) \quad (13)$$

where LambertW is a function which satisfies LambertW($x$) exp( Lambert W($x$) ) = $x$. The solution of

$$\frac{d}{dz}\rho_2 = (-g_B I_{ss}\mu_2 + \alpha)(1-\rho_2)$$
$$\frac{d}{dz}\mu_2 = -(g_B I_{pp}\rho_2 + \alpha)(1-\mu_2) \quad (14)$$

$$\mu_2(z) \simeq \frac{c_2}{I_{ss}}\left( e^{\alpha(z-L)} - 1 + \frac{2\alpha}{c_2 g_B} \ln\left( \frac{I_{ss}(z)}{I_{ss}(z=\frac{L}{2})} \right) \right) \quad (15)$$

$$\rho_2(z) \simeq 1 - e^{\left( 2\alpha \ln\left( \frac{I_{ss}(z=\frac{L}{2})}{I_{ss}(z=L)} \right) - c_2 g_B - \alpha \right)z + \frac{c_2 g_B}{\alpha} e^{-\alpha L}(e^{\alpha z}-1)} \quad (16)$$

In this case new $c_2$ is inserted Eq.(4) to get new values of $I_{pp}$ and $I_{ss}$ intensities. The evolution of Power and Stokes waves for saturation region is;

$$I_{p(Saturation)} = I_{pp}(1-\rho_2)$$
$$I_{s(Saturation)} = I_{ss}(1-\mu_2) \quad (17)$$

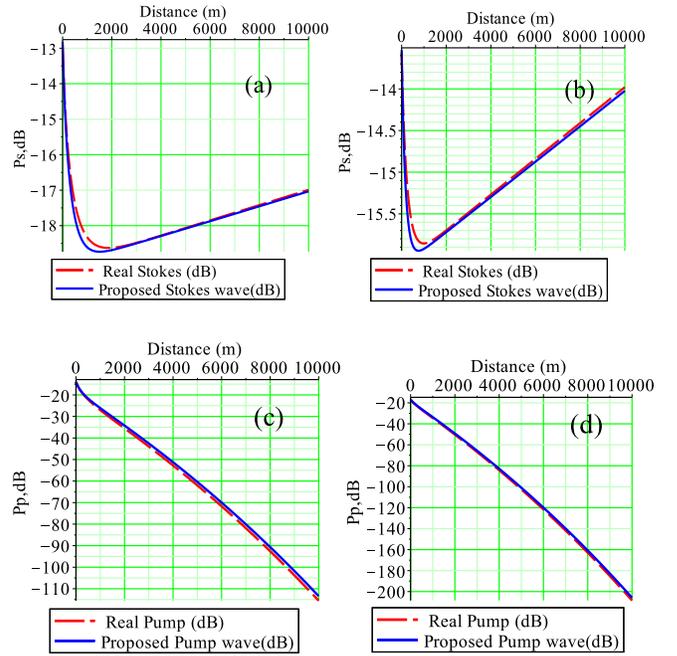

Fig. 4. Stokes wave P$s$ (a), (b) and Pump wave P$p$ (c),(d) versus $z$ . In [(a), (c)], $P_{p0}$=40 mW, $P_{sL}$= 20µW, (k = 55), in [(b), (d)] and $P_{p0}$=20 mW, $P_{sL}$= 40µW ,(k = 28). Thick solid curves, approximate analytical solution (10a) and (10b); dashed curves, full numerical solution of Eq. (1). In both figures, L=10 km, $A_{eff}$ = 80µm², $g_B$=1.091x10⁻¹¹, α=0.215 dB/km.

The BFA gain in saturation can be calculated from (17) and expressed in terms of the physical parameters as

$$G_{BFA} = \frac{A_{eff}}{P_{sL} g_B} \left( \begin{array}{c} -e^{-\alpha L} c_2 g_B + g_B \frac{P_{p0}}{A_{eff}} \\ -2\alpha \ln\left( \frac{\frac{P_{p0}}{A_{eff}} e^{c_2 g_B \frac{L}{2}} - \frac{P_{p0}}{A_{eff}} + c_2}{c_2} \right) \end{array} \right) \quad (18)$$

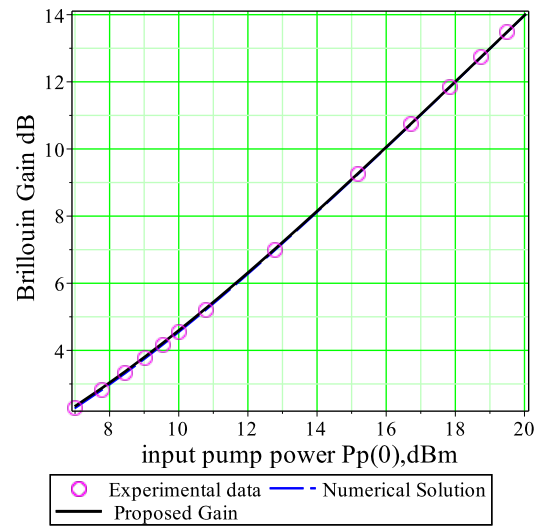

Fig.5. BFA gain versus the input pump power $P_p(0)$ for saturation region. Open circles, experimental data; thick solid curves, predictions

of the analytical formula (18); dashed curves, numerical results. $P_{SL}$ = 4 mW, L=10 km, $A_{eff}$ = 80μm$^2$, $g_B$=1.091x10$^{-11}$, α=0.215 dB/km

The experimental results of the amplifier gain, $G_{BFA}$, and the theoretical predictions show a high-level of agreement (discrepancy 2%) with our analytical formula (18) [Fig. 5]. We note, that Eq. (18) is applicable only when $\Lambda < 0$. Table 1 summarizes the usage of the equations described

**Table 1**
Brief usage of the Equations.

| Criteria | Region | Equation |
|---|---|---|
| $P_{p0} < P_{cr} \approx \left(\Lambda + \frac{A_{eff}}{0.08 \times g_B L}\Lambda\right)$ | UPA | Eqs. 11 |
| $P_{p0} > P_{cr} \approx \left(\Lambda + \frac{A_{eff}}{0.08 \times g_B L}\Lambda\right)$ | High-Gain Region | Eqs. 10 |
| $\Lambda < 0$ | Saturation Region | Eqs. 17 |

Brillouin sensors call for the solution of the steady state equations (Eq.(1a-1b)). This set of equations is valid for pulses larger than phonon lifetime (≈10 ns.), which is equivalent to a spatial resolution, $w > 1$m. Assuming a uniform gain over the spatial resolution, we can integrate it over the range [$z, z+w$]. The variation in the intensity of the cw signal due to the interaction with the pulse at position $z$ is given by [14],

$$G = \frac{I_s(z)}{I_s(z+w)} \quad (19)$$

where $G$ is the Brillouin loss amplitude, which varies with the sensing fiber characteristics as $g(v_B, \Delta v_B)$, $L$, $z$. The two beams involved in the sensing process have a measurable effect through $w$, $I_p$ and $I_s(L)$. To investigate the worst-case analysis we use the metric of the relative error (RE) given as,

$$RE = \left|\left(I_s - I_{s-computed}\right)/I_s\right| \quad (20)$$

Using the RE, we investigate the worst-case analysis for the data from Fig.1 [b-d], i.e., $P_{p0}$=50 mW, $P_{sL}$= 50μW, and find the maximum RE to be 9 % implying that it can be accurately used reliably even with the usage of high pump powers.

## Conclusions

We have presented an approximate analytical solutions to the system of SBS equations in a lossy medium that accurately describes the regime of depleted pump in three different pump-intensity regions, namely undepleted pump approximation, high gain and saturation regions. Three limits of the separation conditions are determined and the respective parameters are obtained. The results obtained can be used to optimize performance of Brillouin fiber amplifiers which are widely used in microwave photonics, radio over fiber technology, and sensing applications which especially uses the high-gain region and spatial resolution $w > 1$m